\documentclass[aps,prb,preprint,showpacs]{revtex4}
\usepackage{natbib}
\usepackage{epsfig}
\usepackage{color}

\begin{document}

\title{Exact results relating spin-orbit interactions in two dimensional 
strongly correlated systems}
\author{N\'ora~Kucska and Zsolt~Gul\'acsi}%$^{a,b}$ and Dieter~Vollhardt$^{a}$}
\affiliation{Department of Theoretical Physics, University of
Debrecen, H-4010 Debrecen, Bem ter 18/B, Hungary}
%$^{(a)}$ Theoretical Physics III, Center for
%Electronic Correlations and Magnetism, Institute for Physics,
%University of Augsburg, D-86135 Augsburg,
%Germany \\
%$^{(b)}$ Department of Theoretical Physics, University of
%Debrecen, H-4010 Debrecen, Hungary}
\date{\today }
%\date{September 24, 2015}

\begin{abstract}
A 2D square, two-bands, strongly correlated and non-integrable system 
is analysed exactly in the presence of many-body spin-orbit interactions via 
the method of Positive Semidefinite Operators. The deduced exact ground states 
in the high concentration limit are strongly entangled, and given by the
spin-orbit coupling are ferromagnetic and present an enhanced carrier mobility,
which substantially differs for different spin projections. The described state
emerges in a restricted parameter space region, which however is clearly 
accessible
experimentally. The exact solutions are provided via the solution of a matching
system of equations containing 74 coupled, non-linear and complex algebraic
equations. In our knowledge, other exact results for 2D interacting systems 
with spin-orbit interactions are not present in the literature.
\end{abstract}

\maketitle

\section{Introduction}

The many-body spin-orbit interaction (SOI) plays an essential role in the
physics of surfaces and interfaces in a continuously increasing number of 
systems of large interest. Indeed, large SOI coupling is present at 
interfaces between heavy
elements (e.g. Pb, Sb, Bi) and non-magnetic materials as Ag, Au, Cu, which can 
lead to surface density waves \cite{I1}, manipulation possibility of spin-orbit
splitting by surface alloying \cite{I2}, or charge- to spin-current 
conversion \cite{I3} with application possibility in 
spintronics; semiconductor reconstructed surfaces using heavy elements as in
the case of Pb on Ge(1,1,1) \cite{I4}, Bi or Tl on Si(1,1,1) \cite{I5},
or Au on Ge(1,1,1) \cite{I6} which can lead to spin control in electronic 
transport applications \cite{I7}; metal-quantum dot configurations influencing
transport properties of Aharonov-Bohm rings \cite{I8}; graphene layers on 
metallic substrates \cite{I9} influencing the transport properties of graphene
\cite{I10}, leads to band splitting and enriched spintronic effects 
\cite{I11} and influences topological insulator properties \cite{I12};
complex oxide interfaces as for example in the case of $LaAlO_3/SrTiO_3$ 
\cite{I13,I14} which even can provide tunable superconductivity at the
interface \cite{I15}; or pnictogen 2D honeycomb type of lattices presenting 
as well magnetic properties \cite{I15a}. Also in other cases
the SOI interaction plays an important role in
the magnetic behaviour of surfaces and interfaces \cite{I16}. Indeed, magnetism
can appear at the interface of two, othervise non-magnetic perovskites 
\cite{I17}; at the interface between $Cr_2O_3$ and overlayers of Pd or Pt
\cite{I18}; Cu or Mn in thin films interfaced with organic molecules
\cite{I19}; nanoparticle surfaces with extremely high surface/volume ratio 
fabricated from otherwise macroscopically non-magnetic materials (e.g. Au, 
or Pd) \cite{I20,I21}; or interfaces in multilayers \cite{I22}. 
In several of these
cases also anomalous magneto-transport measurements have been
reported \cite{I8,I16,I23}.

When SOI is effective at the interface, often it happens that also the inter 
electronic interaction is strong, i.e. the system is strongly correlated. 
This is the case of metal - quantum dot - 
metal configurations \cite{I8}; interfaces present in between complex oxids
or perovskites \cite{I17}; osmate double perovskites \cite{I23a};
GaAs heterostructures \cite{I23}; iridates 
\cite{I24}, iridium based heterostructures \cite{I25} or 
iridates in perovskit- and 
{
honeycomb-based} structures \cite{I26};
and even heterostructures with organic materials \cite{I19}. In these strongly
correlated systems,  the effect of the spin-orbit 
interaction is not yet well understood \cite{I24}, the band splitting (caused
by SOI) in the presence of the inter electronic interaction even considered a
fundamental effect is not understood in details \cite{I23}, the interplay of
strong electron correlation and large SOI is relatively less explored
\cite{I23a}, and  perturbative treatement being inconclusive cannot be applied
\cite{I26}. From the other side is known that SOI has major effects on basic
model results describing strong correlations as in the case of the
Hubbard model \cite{I27}, periodic Anderson model \cite{I28}, or Bethe ansatz
exact solutions derived for integrable cases, which are strongly affected 
\cite{I28a}. Furthermore, often correlations even enhance SOI \cite{I29}, 
or vice versa, SOI provides mechanism for stong correlation effects as e.g. in 
the case of metal-insulator transition \cite{I31}, and that multi-orbital 
treatement is important for the description of effects caused by SOI 
\cite{I29,I30}. Till today such systems have been 
{
analysed} only by exact
diagonalization technique on small samples \cite{I29}, non-equilibrium 
Green-function techniques \cite{I8}, and finally, numerical procedures
e.g.  variational Monte Carlo \cite{I32}, Monte Carlo simulations combined
with spin-wave theory \cite{I33}, or density-functional theory
based approximations \cite{I30}.

Contrary to the importance of this field, exact results relating 2D strongly 
correlated systems containing SOI are not known today. The difficulty of
this job lies in the fact that such systems are non-integrable, and because of
this reason, only special techniques are possible to be applied in order to
deduce exact results. In this paper we begin to fill up this gap by presenting
in our knowledge the first 2D exact ground states for a two band strongly
correlated system containing SOI, using the method of positive semidefinite
operators whose applicability does not depend on dimensionality and 
integrability \cite{T1,T2,T3}, (see also the review in Ref.[\cite{T4}]).  
One notes, that the 
method has been previously applied in conditions unimaginable 
{
before} in the 
context of exact solutions, as disordered systems in 2D [\cite{T5}]; 
multiband systems in 2D [\cite{T6}] and 3D [\cite{T7}];  stripe, 
checkerboard and droplet states in 2D [\cite{T8}]; delocalization 
{
effect} of the
Hubbard repulsion in 2D [\cite{T8a}]; or different non-integrable 
chain structures  \cite{T2,T3,T9,T10,T11}. Here we focus on magnetic properties
and show that given by the interplay of SOI and correlations, ferromagnetism is
possible to be induced on surfaces and interfaces increasing in the same time,
differentiated for different spin projections, the mobility of carriers.

The remaining part of the paper is structured as follows: Section II. describes
the studied system, Section III. presents an insight in the physical behavior
of the system, Section IV. shows the transformation in positive semidefinite 
form of the Hamiltonian, Section V. deduces the exact ground states, 
Section VI. analyses the physical properties of the deduced ground states,
and finally, the Summary and Conclusions in Section VII. closes the 
presentation. The three Appendices A,B,C contain the mathematical details of 
the deductions.

\section{The system analysed}

One analyses a square itinerant system in two dimensions ($2D$)
containing a correlated
band (denoted hereafter by $f$) which experiences the action of the on-site 
Coulomb repulsion $U_f > 0$, and is hybridised with a non-correlated band 
(denoted for simplicity by $d$). The one particle part ($\hat H_{kin}$) of the 
Hamiltonian ($\hat H$) contains besides on-site one-particle potentials 
($\epsilon$), on-site ($V_0$) and nearest-neighbor ($V_{i,j}$)
hybridizations, also nearest-neighbor hopping terms ($t^f_{i,j},t^d_{i,j}$). In
between the hopping terms, given by the many-body spin-orbit interactions, also
spin-flip hopping terms are present. In these conditions, using the
notations $c=d,f$, and ${\bf p}={\bf x},{\bf y}$, where ${\bf x}$ and ${\bf y}$
are representing the Bravais vectors of the $2D$ system, introducing the
local
\begin{eqnarray}
\hat H_{c,0}=\sum_{\bf i} \sum_{\alpha=\uparrow,\downarrow} \epsilon_c^{\alpha,\alpha}
\hat c^{\dagger}_{{\bf i},\alpha} \hat c_{{\bf i},\alpha}, \quad
\hat V_0 =\sum_{\bf i}[(\sum_{\alpha=\uparrow,\downarrow}V_0^{d,f,\alpha,\alpha}
\hat d^{\dagger}_{{\bf i},\alpha} \hat f_{{\bf i},\alpha})+ H.c.],
\label{EQ1}
\end{eqnarray}
and nearest-neighbor
\begin{eqnarray}
&&\hat H_{c,{\bf p}}= \sum_{\bf i} [(\sum_{\alpha=\uparrow,\downarrow} t^{c,\uparrow,\uparrow}_{
\bf p} \hat c^{\dagger}_{{\bf i}+{\bf p},\alpha} \hat c_{{\bf i},\alpha}) + t^{c,\downarrow,
\uparrow}_{\bf p} \hat c^{\dagger}_{{\bf i}+{\bf p},\downarrow} \hat c_{{\bf i},\uparrow} +
t^{c,\uparrow,\downarrow}_{\bf p} \hat c^{\dagger}_{{\bf i}+{\bf p},\uparrow} 
\hat c_{{\bf i},\downarrow} + H.c.],
\nonumber\\
&&\hat V_{\bf p}=\sum_{\bf i} [\sum_{\alpha=\uparrow,\downarrow}(V^{d,f,\alpha,\alpha}_{\bf p}
\hat d^{\dagger}_{{\bf i}+{\bf p},\alpha} \hat f_{{\bf i},\alpha} + V^{f,d,\alpha,\alpha}_{\bf p}
\hat f^{\dagger}_{{\bf i}+{\bf p},\alpha} \hat d_{{\bf i},\alpha}) + H.c.],
\label{EQ2}
\end{eqnarray}  
one-particle contributions, one obtains for $\hat H = \hat H_{kin}+\hat H_{int}$
the expressions
\begin{eqnarray}
&&\hat H_{kin}=\hat V_0 + \sum_{{\bf p}={\bf x},{\bf y}} [ \hat V_{\bf p} +
\sum_{c=d,f}(\hat H_{c,0}+\hat H_{c,{\bf p}}) ],
\nonumber\\
&&\hat H_{int}= \sum_{\bf i} U_f \hat n^f_{{\bf i},\uparrow}\hat n^f_{{\bf i},\downarrow},
\label{EQ3}
\end{eqnarray}
where $\hat c_{{\bf j},\alpha}$, for $c=d,f$ are canonical Fermi operators, and 
$\hat n^f_{{\bf i},\alpha}=\hat f^{\dagger}_{{\bf i},\alpha} \hat f_{{\bf i},\alpha}$.
As seen from the last row of Eq.(\ref{EQ3}), the inter-electronic interaction 
term is represented by the Hubbard interaction in the correlated band. As was
mentioned above, the spin-flip hopping contributions in $\hat H_{c,{\bf p}}$ 
originate from the many-body spin-orbit interactions. Often, these interactions
are taken into account on a phenomenological ground \cite{I27,M2,M3}. 
But one notes that usually such terms emerge via the Rashba and Dresselhaus 
contributions which for 2D square lattice provide
\begin{eqnarray}
\hat H_{SO,c} = \sum_{\bf i} [\sum_{{\bf p}={\bf x},{\bf y}}(V^{c,{\bf p}}_{\uparrow,\downarrow} 
\hat c^{\dagger}_{{\bf i}+{\bf p},\uparrow} \hat c_{{\bf i},\downarrow} +
V^{c,{\bf p}}_{\downarrow,\uparrow} \hat c^{\dagger}_{{\bf i}+{\bf p},\downarrow} \hat c_{{\bf i}
,\uparrow}) + H.c.],
\label{EQ4}
\end{eqnarray}
where, for fixed $c$, one has $V^{c,{\bf x}}_{\uparrow,\downarrow}=V^c_R -iV^c_D,
V^{c,{\bf x}}_{\downarrow,\uparrow}=-V^c_R -iV^c_D, V^{c,{\bf y}}_{\uparrow,\downarrow}=V^c_D -
iV^c_R, V^{c,{\bf y}}_{\downarrow,\uparrow}=-V^c_D -iV^c_R$, the notation $V^c_R,V^c_D$ 
signaling the Rashba, and Dresselhaus interaction strengths respectively, 
see [\cite{RD}]. With the contributions presented in Eq.(\ref{EQ4}), one has
e.g. $t^{c,\uparrow,\downarrow}_{\bf p}=V^{c,{\bf p}}_{\uparrow,\downarrow}$, etc.

As we mentioned in the Introduction, the SOI effects in the presence of
strong correlations must be treated at multibands level. Exactly for this
reason, we present our study for a two-band type of structure. On this line
we note that the presence of two bands does not diminish the applicability of
the deduced results because for a multiband material the theoretical 
description is given usually by projecting the multiband structure in a 
few-band picture \cite{RDA}, which is stopped here only for its relative 
simplicity at two-bands level.

\section{Insight in the physical behavior of the system}

In order to present an insight in the physical behavior of the system
let us concentrate on the bare band structure (see Appendix B)
provided by $\hat H_{kin}$. For simplicity we exemplify by using ${\bf k}_y=0$
plots, the first Brillouin zone in ${\bf k}_x$ direction being placed in 
between the dashed lines (note that all directions present similar qualitative
behavior). Fig.1a shows a pedagogical
band structure without spin-flip 
terms i.e. without spin-orbit interactions. As it can be observed, 
each energy band is double degenerated containing both spin projections.  
Fig.1b exemplifies what is happening when spin-flip contributions are turned on:
the double spin-projection degeneracy is lifted. The resulting $\pm$ bands are
no more spin-projection degenerated. This notation underlines that now
the resulting non-degenerate bands contain usually all spin projections but 
with different weights.

%%%%%%%%%%%%%%%%%%%%%%%%%%%%%%%%%%%%%%%%%%%%%%%%%%%%%%%%%%%%%%%%%%%%%%%%%%%%%
% FIGURE 1
%%%%%%%%%%%%%%%%%%%%%%%%%%%%%%%%%%%%%%%%%%%%%%%%%%%%%%%%%%%%%%%%%%%%%%%%%%%%%
\begin{figure}[h]
\centerline{\includegraphics[width=8cm,height=4cm]{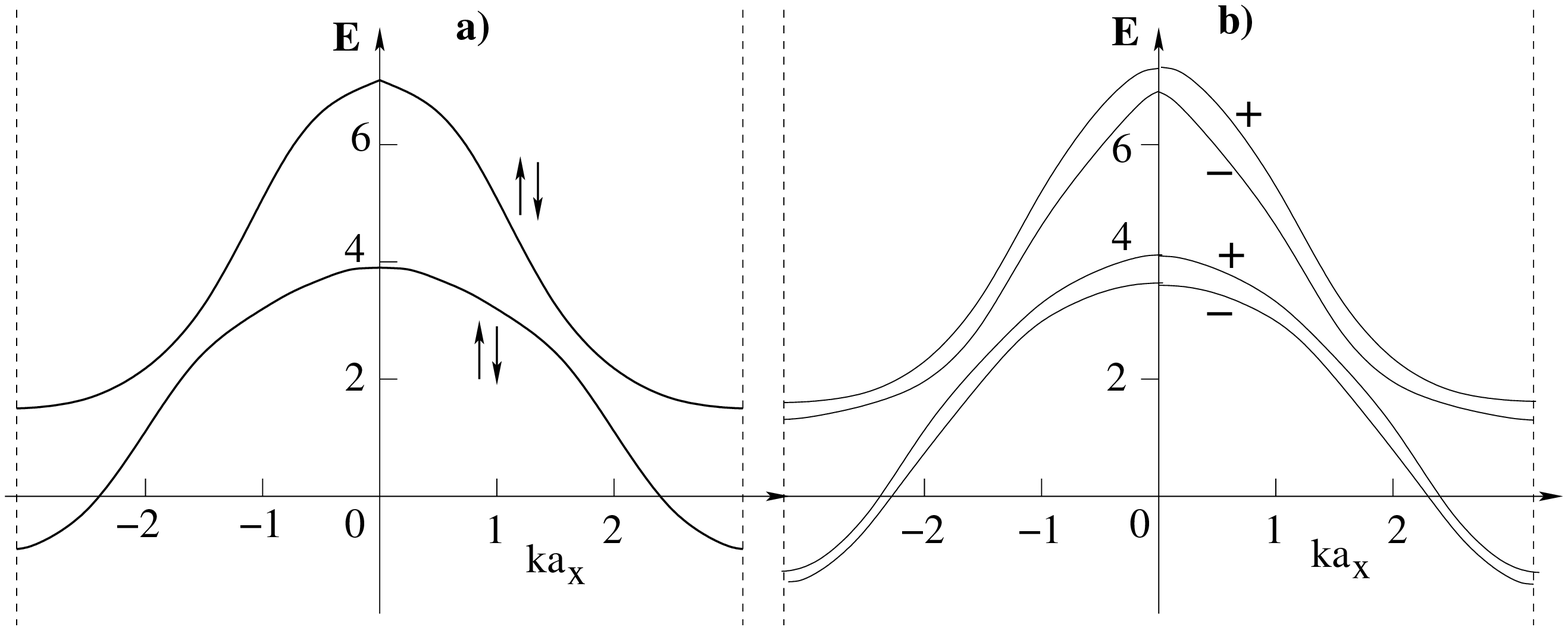}} 
%\centerline{\epsfbox{NoraBandsFig1u.eps}}
\caption{a) The band structure without spin-flip contributions. All parameters 
are expressed in $t^{d,\sigma,\sigma}_x=t^d_x$ units and not depend on the 
$\sigma$ value (i.e. the parameter values for $\uparrow,\uparrow$, and
$\downarrow,\downarrow$ indices are equal). In the provided example, 
non-zero parameters are
$t_x^{d,\uparrow,\uparrow}=1.0;t_y^{d,\uparrow,\uparrow} =1.1;
\epsilon_d^{\uparrow,\uparrow}=0.5; t_x^{f,\uparrow,\uparrow}=1.5; 
t_y^{f,\uparrow,\uparrow}=1.6; \epsilon_f^{\uparrow,\uparrow}=0.1;
V_0^{d,f,\uparrow,\uparrow}=-1.2; V_x^{d,f,\uparrow,\uparrow}=V_x^{f,d,\uparrow,\uparrow}=0.6; 
V_y^{d,f,\uparrow,\uparrow}=V_y^{f,d,\uparrow,\uparrow}=0.7$. b) Spin-flip terms are 
turned on. In this example, besides the a) non-zero parameters, one also has  
$\epsilon_d^{\uparrow,\downarrow}=0.2; \epsilon_f^{\uparrow,\downarrow}=0.25$. 
For the $\pm$ band notation see text.} 
\label{fig1}
\end{figure}
%%%%%%%%%%%%%%%%%%%%%%%%%%%%%%%%%%%%%%%%%%%%%%%%%%%%%%%%%%%%%%%%%%%%%%%%%%%%%

Following the pedagogical example from Fig.1, the plot presented in Fig.2
shows a band structure emerging together with the spin-orbit contributions
present in the Hamiltonian, containing the
imput from (\ref{EQ4}) and leading to $\epsilon_{d,{\bf k}}^{\uparrow,\downarrow}=2
t_x^{d,\uparrow,\downarrow} ( i \sin {\bf k}{\bf x} + e^{i\chi} \cos {\bf k}{\bf y}), 
\epsilon_{f,{\bf k}}^{\uparrow,\downarrow}=2t_x^{f,\uparrow,\downarrow} ( i \sin {\bf k}
{\bf x} + e^{i\chi} \cos {\bf k}{\bf y})$ defined for the band structure
calculation in Appendix B. As seen in Fig.2, the spin projection degeneracy,
as before is lifted, the motivation for the $\pm$ notation is as 
explained for Fig.1. 
  
%%%%%%%%%%%%%%%%%%%%%%%%%%%%%%%%%%%%%%%%%%%%%%%%%%%%%%%%%%%%%%%%%%%%%%%%%%%%%
% FIGURE 2
%%%%%%%%%%%%%%%%%%%%%%%%%%%%%%%%%%%%%%%%%%%%%%%%%%%%%%%%%%%%%%%%%%%%%%%%%%%%%
\begin{figure}[h]
\centerline{\includegraphics[width=6cm,height=6cm]{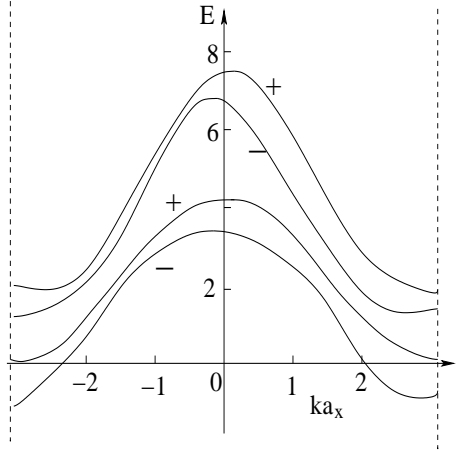}} 
%\centerline{\epsfbox{QuadriUj.eps}}
\caption{Band structure in the presence of spin-orbit interaction considered
in the present paper, see text. The nonzero $\hat H_{kin}$ parameters are as in
Fig.1a, besides which also $t_x^{d,\uparrow,\downarrow}=0.1,
t_x^{f,\uparrow,\downarrow}=0.25, \chi=\pi/2$ are considered. The energy and all
$\hat H_{kin}$ parameters are expressed in $t^d_x$ units. Note that ${\bf k}\cdot {\bf a}_{\bf y} \ne 0$ holds.} 
\label{fig2}
\end{figure}
%%%%%%%%%%%%%%%%%%%%%%%%%%%%%%%%%%%%%%%%%%%%%%%%%%%%%%%%%%%%%%%%%%%%%%%%%%%%% 

This lifted spin projection degeneracy produces the peculiarities measured 
on surfaces in the presence of the spin-orbit interactions which are seen 
as well e.g. in magnetotransport \cite{I3,I17}, or magnetic behavior 
\cite{I16,I19,I22,M7}.

The presented observations motivated our efforts to analyze in exact 
terms the effects of spin-orbit interactions on surfaces: 
can or not such contribution, in the presence of inter-electronic interactions
produce surface ferromagnetism in multiband systems which are often 
{
strongly correlated?} 
The exact study is necessary because i) spin-orbit contributions 
are usually relatively small in comparison to other system parameters emerging 
in strongly correlated systems, and ii) as presented in the introduction,
low or moderate order approximations 
{
do not provide} reliable information in this
field. The results of our study are presented below providing a positive answer
to the above formulated question. 

\section{The Hamiltonian transformed in positive 
semidefinite form}

Since the analysed system is strongly correlated and our aim is to provide
essential non-altered information for it, we deduce exact results in our study.
Taking into account that the model used for description is a 2D non-integrable
model, special techniques must be used for this purpose. Because of this
reason, as we mentioned previously, one uses below a technique based on 
positive semidefinite operator properties.

The first step of the technique transforms in exact terms the system Hamiltonian
in a positive semidefinite form
\begin{eqnarray}
\hat H = \hat P +C
\label{EQ5}
\end{eqnarray}
where $\hat P$ is a positive semidefinite operator and $C$ is a scalar. In the 
present case, since $\hat H$ in (\ref{EQ3}) contains spin-flip terms as well,
one uses for this transformation for the first time block operators that mix 
the spin indices. For each unit cell defined at the lattice site ${\bf i}$, 
one introduces two block operators $\hat A_{\bf i}$ and $\hat B_{\bf i}$, denoted
for simplicity by $G=A,B$, where one has
\begin{eqnarray}
\hat G_{\bf i} = \sum_{c=d,f}\sum_{\alpha=\uparrow,\downarrow} 
(g_{G,c,1,\alpha} \hat c_{{\bf i},\alpha} + g_{G,c,2,\alpha}\hat c_{{\bf i}+{\bf x},\alpha} +
g_{G,c,3,\alpha} \hat c_{{\bf i}+{\bf x}+{\bf y},\alpha} + g_{G,c,4,\alpha}\hat c_{{\bf i}+
{\bf y},\alpha}),
\label{EQ6}
\end{eqnarray}
where $g_{G,c,n,\alpha}$ for each fixed $G$ represent 16 numerical prefactors
(i.e. $c=d,f$, $\alpha=\uparrow,\downarrow$, $n=1,2,3,4$)
which, given by the Bravais translational symmetry of the system, are the same
in each cell defined at arbitrary site ${\bf i}$. In fact the block operators
$\hat G_{\bf i}$ are linear combinations of fermionic annihilation operators 
$\hat c_{{\bf j},\alpha}$, for all $c=d,f$ and $\alpha=\uparrow,\downarrow$, acting
on the four sites of the unit cell defined at the lattice site ${\bf i}$ 
containing the four sites ${\bf i}, {\bf i}+{\bf x}, {\bf i}+{\bf x}+{\bf y}, 
{\bf i}+{\bf y}$, numbered (anti clockwise starting from ${\bf i}$) by the 
in-cell site index $n=1,2,3,4$ (see Fig.1).

%%%%%%%%%%%%%%%%%%%%%%%%%%%%%%%%%%%%%%%%%%%%%%%%%%%%%%%%%%%%%%%%%%%%%%%%%%%%%
% FIGURE 3
%%%%%%%%%%%%%%%%%%%%%%%%%%%%%%%%%%%%%%%%%%%%%%%%%%%%%%%%%%%%%%%%%%%%%%%%%%%%%
\begin{figure}[h]
\centerline{\includegraphics[width=4cm,height=4.5cm]{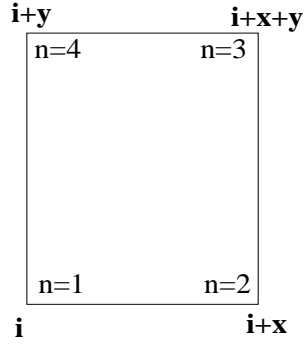}} 
%\centerline{\epsfbox{QuadriUj.eps}}
\caption{Unit cell defined at the lattice site ${\bf i}$ with in-cell notations
of sites $n=1,2,3,4$. The Bravais vectors of the 2D lattice are denoted by 
${\bf x}$ and ${\bf y}$.} 
\label{fig3}
\end{figure}
%%%%%%%%%%%%%%%%%%%%%%%%%%%%%%%%%%%%%%%%%%%%%%%%%%%%%%%%%%%%%%%%%%%%%%%%%%%%%

Using the introduced block operators, the Hamiltonian transformed in positive
semidefinite form (\ref{EQ5}) becomes
\begin{eqnarray}
&&\hat P = \hat P_G + \hat P_1, \: \:
\hat P_G = \sum_{\bf i} \sum_{G=A,B} \hat G_{\bf i} \hat G^{\dagger}_{\bf i}, \: \:
\hat P_1 = U_f \sum_{\bf i}  \hat P_{\bf i}, 
\nonumber\\
&& C=\eta N - U_fN_{sit} - \sum_{G=A,B} \sum_{\bf i} z^{G}_{\bf i}, \: \: U_f > 0,
\label{EQ7}
\end{eqnarray}
where $N$ represents the number of electrons, $N_{sit}$ gives
the number of lattice 
sites, $z^G_{\bf i}=\{ \hat G_{\bf i}, \hat G^{\dagger}_{\bf i} \}$, $\eta$ is a 
numerical parameter determined by the matching equations [see (\ref{A23})], 
while the positive
semidefinite operator $\hat P_{\bf i}= \hat n^f_{{\bf i},\uparrow} \hat n^f_{{\bf i},
\uparrow} -(\hat n^f_{{\bf i},\uparrow} +\hat n^f_{{\bf i},\uparrow}) +1$
attains its minimum eigenvalue zero when at least one electron is present on
the site ${\bf i}$. One notes that in obtaining (\ref{EQ7}), periodic boundary
conditions have been used in both directions.

The matching equations corresponding to the transformation (\ref{EQ5},\ref{EQ7})
(since are representing a coupled non-linear system of 74 equations) are 
presented together with their solution in Appendix A. These equations provide
the block operator parameters $g_{G,c,n,\alpha}$ and the prefactor $\eta$ 
[see (\ref{EQ7})] in 
function of the Hamiltonian parameters present in (\ref{EQ3}) (i.e. 
$t^{c,\alpha,\alpha'}_{\bf p}, V^{c,c',\alpha,\alpha'}_{\bf p}, \epsilon^{\alpha,\alpha'}_c,
V^{c,c',\alpha,\alpha'}_0, U_f$). One needs two block operators $G=A,B$ in order to
cancel out from (\ref{EQ7}) the contributions not present in the starting
$\hat H$ described in (\ref{EQ3}). The matching equations are obtained by 
effectuating the operations presented in the right side of (\ref{EQ7}) and 
equating each obtained term with a given operator structure with the same
operator term present in the starting Hamiltonian (\ref{EQ3}). For example, from
the first row of (\ref{EQ7}), for the hopping term $\hat d^{\dagger}_{{\bf i}+
{\bf x}+{\bf y},\uparrow,} \hat d_{{\bf i},\uparrow,}$ we obtain the coefficient
$-(g^*_{A,d,3,\uparrow}g_{A,d,1,\uparrow} +  g^*_{B,d,3,\uparrow}g_{B,d,1,\uparrow})$. But the
hopping term $\hat d^{\dagger}_{{\bf i}+{\bf x}+{\bf y},\uparrow,} \hat d_{{\bf i},\uparrow,}$,
being cell-diagonal (i.e. next nearest-neighbor 
{
hopping}), is not present in the
starting $\hat H$ from (\ref{EQ3}). Hence the matching equation corresponding
to this hopping term becomes 
$g^*_{A,d,3,\uparrow}g_{A,d,1,\uparrow} +  g^*_{B,d,3,\uparrow}g_{D,d,1,\uparrow}=0$. 
Similarly, for the hopping term  $\hat d^{\dagger}_{{\bf i}+
{\bf x},\uparrow,} \hat d_{{\bf i},\uparrow,}$ from the first row of (\ref{EQ7})
one obtains the coefficient $-(g^*_{A,d,2,\uparrow}g_{A,d,1,\uparrow} +
g^*_{A,d,3,\uparrow}g_{A,d,4,\uparrow} +g^*_{B,d,2,\uparrow}g_{B,d,1,\uparrow} +
g^*_{B,d,3,\uparrow}g_{B,d,4,\uparrow})$. In the starting Hamiltonian (\ref{EQ3}),
the $\hat d^{\dagger}_{{\bf i}+{\bf x},\uparrow,} \hat d_{{\bf i},\uparrow,}$ hopping term
has the coefficient $t^{d,\uparrow,\uparrow}_{\bf x}$, hence the corresponding
matching equation becomes $-t^{d,\uparrow,\uparrow}_{\bf x}=
g^*_{A,d,2,\uparrow}g_{A,d,1,\uparrow} + g^*_{A,d,3,\uparrow}g_{A,d,4,\uparrow} +
g^*_{B,d,2,\uparrow}g_{B,d,1,\uparrow} + g^*_{B,d,3,\uparrow}g_{B,d,4,\uparrow}$. All matching
equations presented in details (together with their solution) in Appendix A,
have been obtained in the same fashion.

\section{The deduced exact ground state wave functions}

Since the lowest possible eigenvalue of a positive semidefinite operator is
zero, once one has the Hamiltonian written in the positive semidefinite form
(\ref{EQ5}), the exact ground state corresponding to (\ref{EQ5}) can be 
obtained in a second step, by constructing the most general wave vector 
$|\Psi_g\rangle$ which 
satisfies the relation $\hat P |\Psi_g\rangle=0$. Several techniques have
been worked out for this purpose \cite{T4}. In the present case, the
ground state wave vector, in the unnormalized form, has the expression
\begin{eqnarray}
|\Psi_g\rangle = \prod_{\bf i} [  (\prod_{G=A,B} \hat G^{\dagger}_{\bf i}) \: 
\hat D^{\dagger}_{\bf i}] |0\rangle, \: \:
\hat D^{\dagger}_{\bf i}= (\gamma_{\uparrow} \hat f^{\dagger}_{{\bf i},\uparrow} + 
\gamma_{\downarrow} \hat f^{\dagger}_{{\bf i},\downarrow}),
\label{EQ8}
\end{eqnarray}
where $\gamma_{\alpha}$, $\alpha=\uparrow,\downarrow$ are numerical prefactors, 
$\prod_{\bf i}$ extends over all
$N_{sit}$ lattice sites, and $|0\rangle$ is the bare 
vacuum with no fermions present.
Note that (\ref{EQ8}) corresponds to $N=3 N_{sit} = (3/4) N_{tot}$ electrons in
the system, where $N_{tot}= 4 N_{sit}$ represents the electron number at complete
system filling. The block operator parameters of $\hat G^{\dagger}_{\bf i}$ 
operators are obtained as solutions of the matching equations and are
explicitly deduced and presented in Appendix A.

The $|\Psi_g\rangle$ wave vector represents the ground state for the following
reasons: i) The $\hat G^{\dagger}_{\bf i}$ operators are linear combinations
of canonical Fermi operators acting on the sites of a finite block, hence 
the equality $\hat G^{\dagger}_{\bf i} \hat G^{\dagger}_{\bf i} =0$ holds. Consequently 
$\hat P_G |\Psi_g\rangle = 0$ is automatically satisfied. Furthermore, ii) the
$\hat P_1$ operator attains its minimum eigenvalue zero when at least one
f-electron is present on all lattice sites ${\bf i}$. But $\prod_{\bf i} \hat D^{
\dagger}_{\bf i}$ introduces at least one f-electron on all sites. As a 
consequence, $\hat P_1 |\Psi_g\rangle = 0$ also holds, hence for
$\hat P=\hat P_G + \hat P_1$ the requirement $\hat P |\Psi_g\rangle = 0$ is
satisfied. We note that the uniqueness of $|\Psi_g\rangle$ from (\ref{EQ8})
can also be demonstrated on the line of Appendix B from
Ref.[\cite{T4}].

We underline that (\ref{EQ8}) represents a highly entangled many-body
ground state. Indeed, if the
products are effectuated from its expression, $|\Psi_g\rangle$ becomes to be a
huge sum over many orthogonal contributions. Furthermore, if at least one 
individual operator connected to an arbitrary ${\bf i}$ site is missing from
the $|\Psi_g\rangle$ expression, (\ref{EQ8}) is no more the
ground state of the studied Hamiltonian. Finally, individual contributions taken
from the right side of (\ref{EQ8}) are not representing one particle 
eigenstates, hence the strongly entangled  many-body nature of 
$|\Psi_g\rangle$ is clearly visible from its expression.

One notes that the ground state (\ref{EQ8}) can also be extended above 
$N^*=3N_{sit}$ to total particle number $N=N^*+N_1$, where $N_1 < N_{sit}$.
Indeed one has
\begin{eqnarray}
|\Psi_g(N>N^*)=\prod_{\bf i} [  (\prod_{G=A,B} \hat G^{\dagger}_{\bf i}) \: 
\hat D^{\dagger}_{\bf i}] \hat F^{\dagger}|0\rangle, \: \:
\hat F^{\dagger}= \prod_{\delta=1}^{N_1} \hat c^{\dagger}_{\delta, 
{\bf k}_{\delta},\sigma_{\delta}},
\label{EQ9}
\end{eqnarray}
where $c_{\delta}$ can be arbitrarily $d,f$; $\sigma_{\delta}$ is an arbitrary
spin projection, and ${\bf k}_{\delta}$ is an arbitrary
momentum from the first Brillouin zone. This is because for $N>N^*$, the 
conditions i) and ii) described below (\ref{EQ8}) are both satisfied, 
(\ref{EQ9}) remaining as well a strongly entangled ground state.

\section{The ground state physical properties}

Once the exact ground state is known, the third step of the method follows,
namely the deduction of the physical properties of the ground state. This is
done by calculating ground state expectation values for quantities of interest.
In this section we deduce in details ground state expectation values at
$N=N^*$ number of particles using the ground state (\ref{EQ8}). The calculations
are done in ${\bf k}$ space representation (see for details Appendix C).

\subsection{The total $S^z$ ground state expectation value}

First, in order to test magnetization properties we calculate the total 
$\hat S^z$ ground state expectation value, where
\begin{eqnarray}
\hat S^z = \sum_{\bf k} \sum_{c=d,f} \hat S^z_{c,{\bf k}}, \: \:
\hat S^z_{c,{\bf k}}=\frac{1}{2}(\hat n^c_{{\bf k},\uparrow}-\hat n^c_{{\bf k},
\downarrow}),
\label{EQ10}
\end{eqnarray}
%%%%%%%%%%%%%%%%%%%%%%%%%%%%%%%%%%%%%%%%%%%%%%%%%%%%%%%%%%%%%%%%%%%%%%%%%%%%%
% FIGURE 4
%%%%%%%%%%%%%%%%%%%%%%%%%%%%%%%%%%%%%%%%%%%%%%%%%%%%%%%%%%%%%%%%%%%%%%%%%%%%%
\begin{figure}[h]
\centerline{\includegraphics[width=5.5cm,height=5cm]{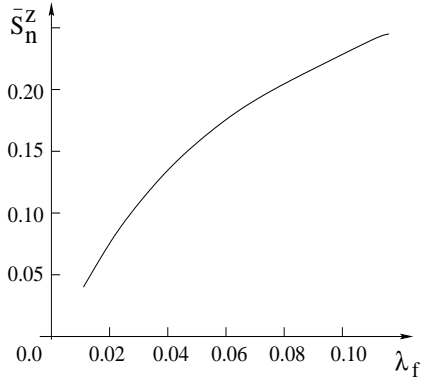}} 
%\centerline{\epsfbox{QuadriUj.eps}}
\caption{The normalized total $\hat S^z$ ground state expectation value
$\bar S^z_n =\bar S^z/\bar S^z_{Max}$ in function of the effective spin-orbit
interaction $\lambda_f=(|V^f_R|^2+|V^f_D|^2)^{1/2}$ in the correlated f-band, 
where $\lambda_f$ is given in $t^{f,\sigma,\sigma}_x$ units.
Note that $\bar S^z_{Max}=1/2$ relation holds.} 
\label{fig4}
\end{figure}
%%%%%%%%%%%%%%%%%%%%%%%%%%%%%%%%%%%%%%%%%%%%%%%%%%%%%%%%%%%%%%%%%%%%%%%%%%%%% 

%%%%%%%%%%%%%%%%%%%%%%%%%%%%%%%%%%%%%%%%%%%%%%%%%%%%%%%%%%%%%%%%%%%%%%%%%%%%%
% FIGURE 5
%%%%%%%%%%%%%%%%%%%%%%%%%%%%%%%%%%%%%%%%%%%%%%%%%%%%%%%%%%%%%%%%%%%%%%%%%%%%%
\begin{figure}[h]
\centerline{\includegraphics[width=12cm,height=5cm]{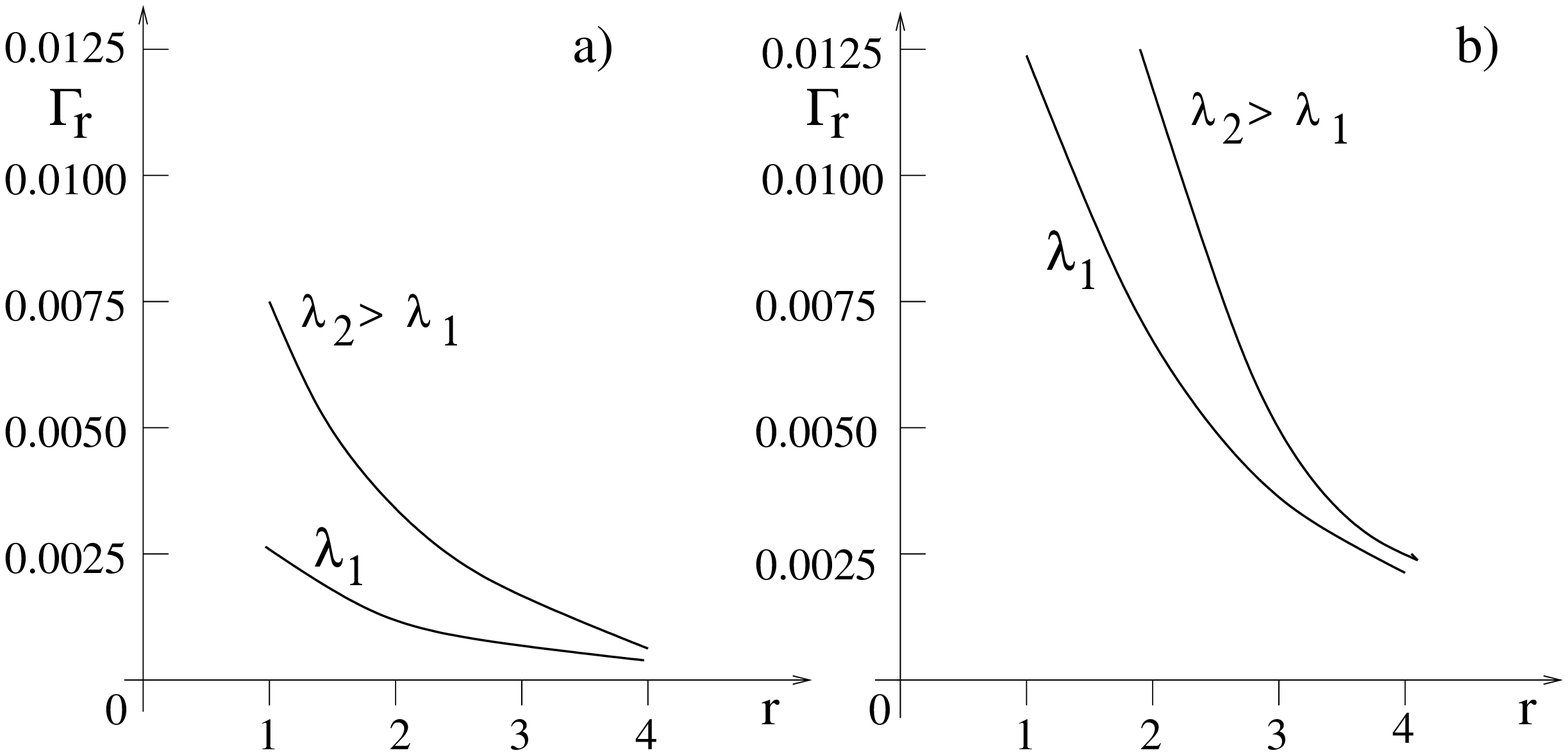}} 
%\centerline{\epsfbox{QuadriUj.eps}}
\caption{The ${\bf r}$ dependent hopping ground state expectation value
$\Gamma_{\bf r}$ in function of the effective total spin-orbit interaction 
$\lambda = (1/2)\sum_{c=d,f} \lambda_c$ exemplified for d electrons in 
${\bf x}$ direction. The distance ${\bf r}$ is given in lattice constant 
units. In the plot one has in $t^{f,\sigma,\sigma}_x$ units $\lambda_1=0.0185$, 
and $\lambda_2=0.0495$, respectively. Note that if $\lambda$ increases, 
$\Gamma_{\bf r}$ increases as well and one has a strong spin projection
dependence. The presented cases: a) negative spin projection to the Z axis, 
and b) positive spin projection to the Z axis.} 
\label{fig5}
\end{figure}
%%%%%%%%%%%%%%%%%%%%%%%%%%%%%%%%%%%%%%%%%%%%%%%%%%%%%%%%%%%%%%%%%%%%%%%%%%%%% 
where the sum over ${\bf k}$ runs over the first Brillouin zone.
Furthermore the expectation values for an arbitrary operator $\hat X$
are standardly deduced via
$\langle \hat X \rangle = \langle \Psi_g| \hat X |\Psi_g\rangle/
\langle \Psi_g| \Psi_g\rangle$. Calculation details are presented in
Appendix C: the norm $\langle \Psi_g| \Psi_g\rangle$ is present in (\ref{EC4}),
the $\langle \hat S^z \rangle/N_{sit}$ ratio, whose magnitude is denoted by 
$\bar S^z$ in Fig.4,  is given in (\ref{EC6}). 
One finds for $\bar S^z$ the result presented in Fig.4.
As can be seen, the ground state expectation value of the total $\hat S^z$
spin increases with the effective spin-orbit interaction
$\lambda_c=(|V^c_R|^2+|V^c_D|^2)^{1/2}$ in the correlated i.e. c=f band. 
The normalized value $\bar S^z_n$ continuously decreases with $\lambda_f$
within the parameter space region in which (\ref{EQ8}) is valid, signaling
that $\langle \hat S^z \rangle$ is non-zero given by the effective spin-orbit
interaction in the correlated band \cite{Note1}. Furthermore since 
$\langle \hat S^z \rangle \ne 0$ automatically implies 
$\langle \hat S^2 \rangle \ne 0$ where $\hat S^2$ is the square of the total
spin, it results that the deduced ground state is ferromagnetic.

\subsection{The ${\bf r}$ dependent hopping ground state 
expectation value}

Furthermore one defines the ${\bf r}$ dependent long-range hopping operator as
\begin{eqnarray}
\hat \Gamma_{{\bf r},c,\sigma} = \frac{1}{N_{sit}}\sum_{\bf j} (\hat c^{\dagger}_{
{\bf j},\sigma} \hat c_{{\bf j}+{\bf r},\sigma} + \hat c^{\dagger}_{{\bf j}+{\bf r},\sigma} 
\hat c_{{\bf j},\sigma}),
\label{EQ11}
\end{eqnarray} 
where, on the background of periodic boundary conditions in both directions,
$\sum_{\bf j}$ runs over all sites, $c=d,f$, and $\sigma$ is an arbitrary spin 
projection. The calculations again are performed in ${\bf k}$ space, and 
calculation details are also contained in Appendix C: the ${\bf k}$ space
expression of (\ref{EQ11}) is given in (\ref{EC7}), while its ground state
expectation value is exemplified in (\ref{EC8}).
The results are presented in Fig. 5. which exemplifies the d-electron behaviour
in ${\bf x}$ direction. As seen, $\Gamma_{\bf r}$ increases if the effective
total spin-orbit interaction $\lambda=(1/2) \sum_{c=d,f} \lambda_c$ increases.
Since $\Gamma_{\bf r}$ is related to the hopping probability, the result
presented in Fig.5 shows that the mobility of carriers increases with the 
spin-orbit interaction, and that the hopping probability (hence mobility) for 
the spin projection in the direction of the spontaneous magnetisation is at 
least five times higher than the hopping probability for the opposed spin 
direction. 

\subsection{Observation relating the obtained solution}

First one notes that using the $N > N^*$ ground state (\ref{EQ9}) in deducing
the ground state expectation values, the results presented in this section
remain qualitatively unchanged: $\bar S^z \ne 0$, and with the 
observation that the ${\bf r}$ dependent hopping $\Gamma_{\bf r}$ substantially
increases for higher distances, the information contained in Figs.4-5
qualitatively remains true.

The second aspect which we must mention, is that the obtained exact results
are connected to a non-integrable system, hence as always in this case 
\cite{T4}, they are linked to a restricted region of the parameter space 
built up from the coupling constants and physical parameters of the Hamiltonian.
This is because the matching system of equations providing the explicit 
expression of the block operator parameters, also leads to interconnections in
between Hamiltonian parameters [see Appendix A, e.g. (\ref{A15})]. These 
interconnections define the restricted parameter space region in which the 
solution is valid. Here we would like to present this region.

We have in the model (besides the on-site one particle potentials and local
hybridizations) only nearest-neighbor hoppings and hybridizations in the kinetic
part of the Hamiltonian. Without spin-flip hopping terms (i.e. spin-orbit 
interaction), all Hamiltonian parameters with opposite spin indices are equal
(i.e. $W^{c,c',\alpha,\alpha}_{\bf p}=W^{c,c',-\alpha,-\alpha}_{\bf p}=W^{c,c'}_{\bf p}$, where 
$W=t,V$, so represents hoppings (in this case $c=c'$), or hybridizations 
(in this case $c \ne c'$as well, $c,c'=d,f$), and one has ${\bf p}={\bf x},
{\bf y}, 0$. Furthermore, taking contributions from the two 
bands, $d,f$, relating W one observes that the matching system of equations 
provides the relation 
\begin{eqnarray}
W^{c,c'}_{\bf p}= w^{\delta_{d,c}+\delta_{d,c'}} W^{f,f}_{\bf p},
\label{EQ12}
\end{eqnarray}
where $w=|u/v|$ is a constant. Such type of relations are 
effectively observed in two-band systems \cite{Hota}, hence often used during
theoretical descriptions \cite{T5}, 
{
consequently do not represent} unrealistic 
restrictions. We note that also the second equality from (\ref{A24}) enters 
in the 
{
category} of the relation (\ref{EQ12}), because it can be written as
$k_1 V_{\bf x}^{c=d,c'=f,\alpha,\alpha}= w t_{\bf x}^{f,f,\alpha,-\alpha}$, where the new, 
strongly reductive factor $k_1$ emerges because of the spin change 
$-\alpha \to \alpha$. If for a concrete system under study is necessary to
accomodate  $t_{\bf x}^{f,f,\alpha,-\alpha}$ to this equality, one notes that 
external electrical potential gradient at the surface can be used \cite{I8} to 
modify the effective spin-orbit interaction value.

The following observation relates the spin-flip hopping terms for which the 
matching equations require a relation of the type (\ref{EQ12}), but with minus 
sign [see (\ref{A16})]. We underline, 
{
that these requirements are also 
realistic.} In order to show this we mention that calculating separately the
d and f contributions to the total $\hat S^z$ ground state expectation value,
one finds that $\langle \hat S^z_d \rangle$ and  $\langle \hat S^z_f \rangle$
{
do not coincide in sign}. Since the f-carriers experience the Hubbard repulsion,
they are the {\it heavier} particles while the 
{
d-carriers are the {\it light}}
particles in the system. But one knows that in two band systems the 
light carriers often try
to compensate partially the heavier particle spin moments as one encounters this
effect in the case of the periodic Anderson model as well \cite{mir1,mir2}.
Consequently, since the d and f spin orientations in average are opposed, the
source term of the spin-orbit interaction\cite{RD}
$\vec \sigma \cdot (\nabla V \times \vec p)$, where $\vec \sigma$ is the
spin orientation,  changes sign. This means that the spin-flip hopping terms 
change sign if one changes the particle type index f to d. 

The last restrictive relation is the first equality of (\ref{A24}) which 
provides a requirement for the value of the local hybridization. But one knows
that the local hybridization can be modified by the concentration of the 
impurities on the surface \cite{v0}, hence this requirement can be 
experimentally satisfied. Interestingly, this requirement relating the local
hybridization can be also enrolled on the line of (\ref{EQ12}), since can be 
written as 
$k_2 k_1 V_{{\bf p}'=0}^{c=d,c'=f,\alpha,\alpha}= w t_{{\bf p}={\bf x}}^{f,f,\alpha,-\alpha}$,
where the second correction factor $k_2$ emerges because of the ${\bf p}$ index
change ${\bf x} \to 0$.

Concluding this subsection, as was presented above, the requirements 
characterizing the restricted parameter space region can be experimentally
easily fulfilled.

\section{Summary and Conclusions}

A two-bands system has been analysed on a square 2D Bravais lattice possessing
a correlated band holding on-site Coulomb repulsion, hybridized with a
non-correlated band, in the presence of many-body spin-orbit interactions. 
Even if the system is non-integrable, exact many-body ground states have been 
deduced for it in the high concentration limit using a technique based on 
Positive Semidefinite Operator properties. For this to be possible, block 
operators holding both spin projections have been used, and at the 
transformation of the system Hamiltonian in a positive semidefinite form, 
solution has been provided for the matching system of equations containing 
74 coupled, non-linear complex-algebraic equations. In our knowledge, other 
exact results for interacting many-body 2D systems in the 
presence of spin-orbit interactions are not present in the literature.
The studied ground state being strongly entangled is ferromagnetic and
presents (differentiated for different spin projections) an enhanced 
mobility of carriers. The increasing effective spin-orbit 
interaction in the correlated band increases the magnetisation, while the total
effective spin-orbit interaction from both bands increases the ${\bf r}$
dependent hopping ground state expectation value enhancing the mobility of
carriers. The hopping probability for spin projection in the direction of the
spontaneous magnetisation is at least five times higher than the hopping 
probability for the opposed spin projection.
Based on the deduced results the emergence of ferromagnetic 
interfaces in between non-magnetic but strongly correlated materials could be 
in principle explained based on the presence of the spin-orbit interaction at
the interface. Since strong mobility increase is observed for a given spin
projection, 
{
applications in spintronics become possible.} 

\section*{Acknowledgements}

N. K. acknowledges the support of UNKP-17-2 New National Excellence
Program of the Hungarian Ministry of Human Capacities, while Z. G. 
acknowledges the support of 
{
the project NKFI-128018 of Hungarian funds for 
basic research and of} the Alexander von Humboldt Foundation. 

\appendix

\section{The matching equations}

\subsection{The system of matching equations}

For more visibility one introduces the notations 
$g_{G=A,c,n,\alpha}=a_{n,c,\alpha}$, and $g_{G=B,c,n,\alpha}=b_{n,c,\alpha}$. Using 
these notations, the matching equations become as follows:

\subsubsection{Matching equations from hopping contributions}

The first  32 matching equations are related to the hopping contributions.
In these equations everywhere one has $c=d,f$, and 
$(\alpha,\alpha') = 
(\uparrow,\uparrow); (\downarrow,\downarrow); (\uparrow,\downarrow);
(\downarrow,\uparrow).$

i) The first 8 equations are related to the nearest-neighbor hopping
terms in ${\bf x}$ direction
\begin{eqnarray}
-t^{c,\alpha,\alpha'}_{\bf x} = a^*_{2,c,\alpha}a_{1,c,\alpha'} +
a^*_{3,c,\alpha}a_{4,c,\alpha'} + b^*_{2,c,\alpha}b_{1,c,\alpha'} +
b^*_{3,c,\alpha}b_{4,c,\alpha'}.
\label{A1}
\end{eqnarray}

ii) Similarly, the second 8 equations are related to the nearest-neighbor 
hopping terms in ${\bf y}$ direction
\begin{eqnarray}
-t^{c,\alpha,\alpha'}_{\bf y} = a^*_{4,c,\alpha}a_{1,c,\alpha'} +
a^*_{3,c,\alpha}a_{2,c,\alpha'} + b^*_{4,c,\alpha}b_{1,c,\alpha'} +
b^*_{3,c,\alpha}b_{2,c,\alpha'}.
\label{A2}
\end{eqnarray}

iii) The next nearest-neighbor hoppings (missing from the starting
$\hat H$ from (\ref{EQ3})) in ${\bf y}+{\bf x}$ direction give the
following 8 homogeneous equations
\begin{eqnarray}
-t^{c,\alpha,\alpha'}_{{\bf y}+{\bf x}} = 0 = a^*_{3,c,\alpha}a_{1,c,\alpha'} +
b^*_{3,c,\alpha}b_{1,c,\alpha'}.
\label{A3}
\end{eqnarray}

iv) The last 8 equations from this group are related again to  next 
nearest-neighbor hoppings missing from the starting
$\hat H$ from (\ref{EQ3}). These hoppings are in  ${\bf y}-{\bf x}$
direction and provide again homogeneous equations
\begin{eqnarray}
-t^{c,\alpha,\alpha'}_{{\bf y}-{\bf x}} = 0 = a^*_{4,c,\alpha}a_{2,c,\alpha'} +
b^*_{4,c,\alpha}b_{2,c,\alpha'}.
\label{A4}
\end{eqnarray}

\subsubsection{Matching equations from non-local hybridizations}

The second 32 matching equations are related to the non-local
hybridizations. For all these equations one has $(c,c')=(d,f);(f,d)$,
and $(\alpha,\alpha') = 
(\uparrow,\uparrow); (\downarrow,\downarrow); (\uparrow,\downarrow);
(\downarrow,\uparrow).$

i) The first 8 equations from this group are related to the non-local
hybridizations in ${\bf x}$ direction
\begin{eqnarray}
-V^{c,c',\alpha,\alpha'}_{{\bf x}} = a^*_{2,c,\alpha}a_{1,c',\alpha'} +
a^*_{3,c,\alpha}a_{4,c',\alpha'} + b^*_{2,c,\alpha}b_{1,c',\alpha'}+
b^*_{3,c,\alpha}b_{4,c',\alpha'}. 
\label{A5}
\end{eqnarray}

ii) The second 8 equations from this group are similarly related to the 
non-local hybridizations in ${\bf y}$ direction
\begin{eqnarray}
-V^{c,c',\alpha,\alpha'}_{{\bf y}} = a^*_{4,c,\alpha}a_{1,c',\alpha'} +
a^*_{3,c,\alpha}a_{2,c',\alpha'} + b^*_{4,c,\alpha}b_{1,c',\alpha'}+
b^*_{3,c,\alpha}b_{2,c',\alpha'}. 
\label{A6}
\end{eqnarray}

iii) As in the hopping case, for the next nearest-neighbor hybridizations
in ${\bf y}+{\bf x}$ direction, missing from the starting Hamiltonian 
(\ref{EQ3}) one obtain 8 homogeneous equations
\begin{eqnarray}
-V^{c,c',\alpha,\alpha'}_{{\bf y}+{\bf x}} = 0 = a^*_{3,c,\alpha}a_{1,c',\alpha'} +
b^*_{3,c,\alpha}b_{1,c',\alpha'}. 
\label{A7}
\end{eqnarray} 

iv) Finally, for the next nearest-neighbor hybridizations
in ${\bf y}-{\bf x}$ direction, again missing from the starting 
$\hat H$ in (\ref{EQ3}), we obtain the last 8 homogeneous equations from
this group, namely
\begin{eqnarray}
-V^{c,c',\alpha,\alpha'}_{{\bf y}-{\bf x}} = 0 = a^*_{4,c,\alpha}a_{2,c',\alpha'} +
b^*_{4,c,\alpha}b_{2,c',\alpha'}. 
\label{A8}
\end{eqnarray} 

\subsubsection{Matching equations from local hybridizations}

Since according to the second equality from (\ref{EQ1}) one has
$V_0^{d,f,\alpha,\alpha'} = (V_0^{f,d,\alpha',\alpha})^*$, the local hybridizations
provide only 4 matching equations. These written for $(c,c')=(d,f)$ and
$(\alpha,\alpha') = 
(\uparrow,\uparrow); (\downarrow,\downarrow); (\uparrow,\downarrow);
(\downarrow,\uparrow)$, give the following relation
\begin{eqnarray}
-V^{d,f,\alpha,\alpha'}_{0} = \sum_{n=1}^4 (a^*_{n,d,\alpha}a_{n,f,\alpha'} +
b^*_{n,d,\alpha}b_{n,f,\alpha'}). 
\label{A9}
\end{eqnarray} 

\subsubsection{Matching equations from local one-particle potentials}

In the case of local one-particle potentials, given by (\ref{EQ1})
one has for each $c=d,f$ the equality $\epsilon_c^{\uparrow,\downarrow}=
(\epsilon_c^{\downarrow,\uparrow})^*$, hence for a fixed $c$ index, three
matching equations appear. Consequently, for $\epsilon_c^{\alpha,\alpha'}$ at
$c=d,f$ one has 6 matching
equations, namely, for $\alpha=\alpha'=\uparrow,\downarrow$, in total 4
equations (note that here $U_d =0$)
\begin{eqnarray}
\epsilon_c^{\alpha,\alpha} +U_c -\eta = \sum_{n=1}^4 (a^*_{n,c,\alpha}a_{n,c,\alpha} +
b^*_{n,c,\alpha}b_{n,c,\alpha}), 
\label{A10}
\end{eqnarray} 
while for $\alpha=\uparrow, \alpha'=\downarrow$ and $c=d,f$, in total 
2 equations
\begin{eqnarray}
\epsilon_c^{\uparrow,\downarrow} = \sum_{n=1}^4 (a^*_{n,c,\uparrow}a_{n,c,\downarrow} +
b^*_{n,c,\uparrow}b_{n,c,\downarrow}). 
\label{A11}
\end{eqnarray} 
Note that the total number of matching equations obtained (\ref{A1} - \ref{A11})
is 74.

\subsection{The solution of matching system of equations}

\subsubsection{The first group of 42 equations}

\underline{The next nearest neighbor contributions}

The unknown variables of the matching equations are the $a_{n,c,\alpha},
b_{n,c,\alpha}$ numerical prefactors that must be expressed in function of
the physical parameters of the starting Hamiltonian. 
The solution process in the first step treates the 32 homogeneous equations
(\ref{A3},\ref{A4},\ref{A7},\ref{A8}) which provides the expression of all 
$a_{n,c\alpha}$ coefficients in function of $b_{n,c,\alpha}$ coefficients as follows.
For both indices $\alpha=\uparrow,\downarrow$ , and both $c=d,f$ one has
\begin{eqnarray}
a_{1,c,\alpha}=-\frac{1}{x}b_{1,c,\alpha}, \: a_{2,c,\alpha}=-\frac{1}{v}
b_{2,c,\alpha}, \: a_{3,c,\alpha}=x^*b_{3,c,\alpha}, \:
a_{4,c,\alpha}=v^*b_{4,c,\alpha},  
\label{A12}
\end{eqnarray}
where $x,v$ are at the moment arbitrary ($\ne 0,\infty$) parameters. (\ref{A12})
gives 16 relations based on which, all 32 homogeneous matching equations 
connected to next nearest neighbor hoppings and hybridizations with zero value
are satisfied.

\underline{The spin-flip hybridization contributions}

In the following step one analyzes the group of 10 equations related to 
non-local (8 equations) and local (2 equations) hybridization terms 
containing spin-flip, which are also missing from the starting Hamiltonian. 
In the case of non-local hybridizations, these contributions are present 
in (\ref{A5},\ref{A6}) as $V^{c,c',\uparrow,\downarrow}_{\bf p}, 
V^{c,c',\downarrow,\uparrow}_{\bf p}$, with ${\bf p}={\bf x},{\bf y}$, and 
$(c,c')=(d,f),(f,d)$. In the case of local hybridizations, the discussed
contributions are present in (\ref{A9}) as $V^{d,f,\uparrow,\downarrow}_0$ and
$V^{d,f,\downarrow,\uparrow}_0$. All these equations allow the expression
of $b_{n,d,\alpha}$ prefactors in function of $b_{n,f,\alpha}$ coefficients as follows
\begin{eqnarray}
b_{1,d,\alpha}=\frac{x u_{\alpha}}{v} b^*_{3,f,-\alpha}, \:
b_{2,d,\alpha}=u_{\alpha}b^*_{4,f,-\alpha}, \:
b_{3,d,\alpha}=-\frac{u_{\alpha}}{x^*v} b^*_{1,f,-\alpha}, \:
b_{4,d,\alpha}=-\frac{u_{\alpha}}{|v|^2} b^*_{2,f,-\alpha}.
\label{A13}
\end{eqnarray}
The equalities (\ref{A13}) represent in total 8 relations. Here $u_{\alpha}$, 
$\alpha=\uparrow,\downarrow$ are at the moment arbitrary ($\ne 0, \infty$)
numerical parameters.

At this moment, from the matching relations only 32 equations remain
[8 from (\ref{A1}), 8 from (\ref{A2}), 4 from (\ref{A5}), 4 from (\ref{A6}),
2 from (\ref{A9}), 4 from (\ref{A10}), and 2 from (\ref{A11})].

\subsubsection{The remaining group of 32 equations}

\underline{The 15 interdependent equations}

From the remaining 32 equations, 17 are linearly independent, and 15 are 
dependent on these. In this subsection we analyse these last 15 interdependent
relations. Taking $u=u_{\uparrow}=-u_{\downarrow}$, the 
interdependences provide i) spin projection independence for non-local
hybridization parameters
\begin{eqnarray}
V^{c,c',\alpha,\alpha}_{\bf p}=V^{c,c',-\alpha,-\alpha}_{\bf p},
\label{A14}
\end{eqnarray}
where $(c,c')=(d,f);(f,d)$, and $\alpha=\uparrow,\downarrow$, hence
(\ref{A14}) gives 4 equations; ii) preserves the spin projection independence
possibility in hopping terms without spin-flip and fixes the magnitude ratio
between d and f nearest-neighbor hoppings and d and f on-site one-particle 
potentials 
\begin{eqnarray}
t^{d,\alpha,\alpha}_{\bf p}=K t^{f,\alpha,\alpha}_{\bf p}, \: 
\bar \epsilon_d^{\alpha,\alpha}= K \bar \epsilon_f^{\alpha,\alpha}
\label{A15}
\end{eqnarray}
where $\bar \epsilon_d = \epsilon_d -\eta$,  $\bar \epsilon_f = \epsilon_f
+U_f -\eta$, and $K=|u/v|^2$, ${\bf p}={\bf x},{\bf y}$ [note that (\ref{A15})
means 6 relations], iii) fixes the magnitude ratio of the absolute values of 
d and f spin-flip hoppings to $K$ as
\begin{eqnarray}
t^{d,\alpha,\alpha'}_{\bf p}=-K t^{f,\alpha,\alpha'}_{\bf p},
\label{A16}
\end{eqnarray}
which represent 4 relations, and finally iv) fixes a proportionality in between 
$\epsilon_d^{\uparrow,\downarrow}$ and  $\epsilon_f^{\uparrow,\downarrow}$ (1 relation) 
\begin{eqnarray}
\epsilon_d^{\uparrow,\downarrow}=\frac{u^*_{\uparrow}u_{\downarrow}}{|v|^2}
\epsilon_f^{\uparrow,\downarrow}.
\label{A17}
\end{eqnarray}
The equations (\ref{A14}-\ref{A17}) are the mentioned 15 interdependences. 

\underline{The remaining 17 matching equations}

From the remaining 17 matching equations first $\epsilon_f^{\uparrow,\downarrow}=0$
is taken, since this parameter is missing from the starting Hamiltonian in
(\ref{EQ3}). This provides the possibility to express $b_{n,f,\alpha}$ for $n=3,4$
in function of the coefficients $b_{m,f,\alpha}$ for $m=1,2$ as follows
\begin{eqnarray}
b_{3,f,\uparrow}=-\frac{1}{x^*}b_{1,f,\uparrow}e^{+i\phi_3}, \:
b_{3,f,\downarrow}=\frac{1}{x^*}b_{1,f,\downarrow}e^{+i\phi_3}, \:
b_{4,f,\uparrow}=-\frac{1}{v^*}b_{2,f,\uparrow}e^{+i\phi_4}, \:
b_{4,f,\downarrow}=\frac{1}{v^*}b_{2,f,\uparrow}e^{+i\phi_4},
\label{A18}
\end{eqnarray}
where at the moment $\phi_3,\phi_4$ are arbitrary phases.
Note that at this step 16 matching equations remain. 5 equations from these
give the following equalities
\begin{eqnarray}
&&V_0^{d,f,\downarrow,\downarrow}=V_0^{d,f,\uparrow,\uparrow}, \:
t^{f,\uparrow,\downarrow}_{\bf y}=\frac{v-x}{1+xv^*}\frac{v^*}{v}e^{-i\phi_4}
t^{f,\uparrow,\downarrow}_{\bf x}, \:
t^{f,\downarrow,\uparrow}_{\bf y}=-\frac{v-x}{1+xv^*}\frac{v^*}{v}e^{-i\phi_4}
t^{f,\downarrow,\uparrow}_{\bf x}, 
\nonumber\\
&&(V^{f,d,\uparrow,\uparrow}_{\bf x})^*=e^{2i\theta_1}e^{i(\phi_3-\phi_4)}
V^{d,f,\uparrow,\uparrow}_{\bf x}, \:
(V^{f,d,\uparrow,\uparrow}_{\bf y})^*=-e^{2i\theta_2}e^{i(\phi_3+\phi_4)}
V^{d,f,\uparrow,\uparrow}_{\bf y}, 
\label{A19}
\end{eqnarray}
where $\theta_1$ is the phase of $Q_1=\frac{xv^*}{1+xv^*}$, and $\theta_2$
is the phase of $Q_2=\frac{xv}{x-v}$. The remaining 11 matching equations
are presented below:

\underline{The remaining 11 matching equations}
\begin{eqnarray}
&&-t^{f,\uparrow,\uparrow}_{\bf x}=(1+\frac{1}{xv^*})(b^*_{2,f,\uparrow}b_{1,f,\uparrow}+
e^{i(\phi_4-\phi_3)}b^*_{1,f,\uparrow}b_{2,f,\uparrow}),
\nonumber\\
&&-t^{f,\downarrow,\downarrow}_{\bf x}=(1+\frac{1}{xv^*})(b^*_{2,f,\downarrow}
b_{1,f,\downarrow}+e^{i(\phi_4-\phi_3)}b^*_{1,f,\downarrow}b_{2,f,\downarrow}),
\nonumber\\
&&-t^{f,\uparrow,\downarrow}_{\bf x}=(1+\frac{1}{xv^*})(b^*_{2,f,\uparrow}b_{1,f,\downarrow}-
e^{i(\phi_4-\phi_3)}b^*_{1,f,\uparrow}b_{2,f,\downarrow}),
\nonumber\\
&&-t^{f,\downarrow,\uparrow}_{\bf x}=(1+\frac{1}{xv^*})(b^*_{2,f,\downarrow}b_{1,f,\uparrow}-
e^{i(\phi_4-\phi_3)}b^*_{1,f,\downarrow}b_{2,f,\uparrow}),
\nonumber\\
&&-t^{f,\uparrow,\uparrow}_{\bf y}= \frac{v-x}{xv}e^{-i\phi_4}(b^*_{2,f,\uparrow}
b_{1,f,\uparrow}-e^{i(\phi_4-\phi_3)}b^*_{1,f,\uparrow}b_{2,f,\uparrow}),
\nonumber\\
&&-t^{f,\downarrow,\downarrow}_{\bf y}= \frac{x-v}{xv}e^{-i\phi_4}(b^*_{2,f,\downarrow}
b_{1,f,\downarrow}-e^{i(\phi_4-\phi_3)}b^*_{1,f,\downarrow}b_{2,f,\downarrow}),
\nonumber\\
&&-V^{d,f,\uparrow,\uparrow}_{\bf x}=\frac{1+xv^*}{xv^*}(\frac{u}{v})^*e^{i\phi_4}
(b_{1,f,\uparrow}b_{2,f,\downarrow}+b_{1,f,\downarrow}b_{2,f,\uparrow}),
\nonumber\\
&&-V^{d,f,\uparrow,\uparrow}_{\bf y}=\frac{v-x}{xv}(\frac{u}{v})^*
(b_{1,f,\uparrow}b_{2,f,\downarrow}-b_{1,f,\downarrow}b_{2,f,\uparrow}),
\nonumber\\
&&-V^{d,f,\uparrow,\uparrow}_0=2(\frac{u}{v})^* [(1+\frac{1}{|x|^2})e^{i\phi_3}
b_{1,f,\uparrow}b_{1,f,\downarrow}+(1+\frac{1}{|v|^2})e^{i\phi_4}
b_{2,f,\uparrow}b_{2,f,\downarrow})],
\nonumber\\
&&\bar \epsilon^{\uparrow,\uparrow}_f=2[(1+\frac{1}{|x|^2})|b_{1,f,\uparrow}|^2 +
(1+\frac{1}{|v|^2})|b_{2,f,\uparrow}|^2],
\nonumber\\
&&\bar \epsilon^{\downarrow,\downarrow}_f=2[(1+\frac{1}{|x|^2})|b_{1,f,\downarrow}|^2 +
(1+\frac{1}{|v|^2})|b_{2,f,\downarrow}|^2].
\label{A20}
\end{eqnarray}
Now, since from (\ref{EQ4}) at ${\bf p}={\bf x}$, the spin-flip hopping terms 
must satisfy $t^{c,\downarrow,\uparrow}_{\bf x}= - (t^{c,\uparrow,\downarrow}_{\bf x})^*$,
from the third and fourth equality of (\ref{A20}) one finds $\theta_1=0, \phi_3=
\phi_4$. 
{
After this step one can see} from (\ref{A20}) that 
$t^{f,\uparrow,\uparrow}_{\bf x}
=t^{f,\downarrow,\downarrow}_{\bf x}$, and $t^{f,\uparrow,\uparrow}_{\bf y}=t^{f,\downarrow,
\downarrow}_{\bf y}$ are satisfied by 
\begin{eqnarray}
b_{2,f,\downarrow}= \frac{b^*_{2,f,\uparrow}b_{1,f,\uparrow}}{b^*_{1,f,\downarrow}},
\label{A21}
\end{eqnarray}
and one remains from (\ref{A20}) with 8 matching equations. Given by 
(\ref{EQ4},\ref{A19}) which require the same absolute value for $t^{f,\uparrow,
\downarrow}_y$ and $t^{f,\uparrow,\downarrow}_x$ one must has $(v-x)/(1+xv^*)=z$ with
$|z|=1$, which provide for $v=|v|e^{i\phi_v}$ the expression $x=|x|e^{i\phi_v}$,
where $|x|=(|v|-1)/(|v|+1)$, and $\theta_2=\phi_v$ holds. Now the last two 
equalities from (\ref{A20}) via
$\bar \epsilon^{\uparrow,\uparrow}_f=\bar \epsilon^{\downarrow,\downarrow}_f$, and
the choise $t^{f,\uparrow,\downarrow}_x \ne 0$ give
\begin{eqnarray}
b_{2,f,\uparrow} =\frac{|v|\sqrt{2}}{|(|v|-1)|}b_{1,f,\downarrow}e^{i\gamma}, \: 
b_{2,f,\downarrow} =\frac{|v|\sqrt{2}}{|(|v|-1)|}b_{1,f,\uparrow}e^{-i\gamma},
\label{A22}
\end{eqnarray}
where $\gamma$ is an arbitrary phase, one can express $\eta$
via $\bar \epsilon^{\uparrow,\uparrow}_f$ as
\begin{eqnarray}
\eta=\epsilon_f^{\uparrow,\uparrow} + U_f - 4\frac{(1+|v|^2)}{(|v|-1)^2}
(|b_{1,f,\downarrow}|^2+|b_{1,f,\uparrow}|^2),
\label{A23}
\end{eqnarray}
and $V^{d,f,\uparrow,\uparrow}_0$, and $t^{f,\uparrow,\downarrow}_{\bf x}$ respectively
become
\begin{eqnarray}
&&-V^{d,f,\uparrow,\uparrow}_0=8 e^{i\chi}\frac{u^*}{|v|} \frac{(1+|v|^2)}{(1-|v|)^2}
b_{1,f,\uparrow}b_{1,f,\downarrow},
\nonumber\\
&&-t^{f,\uparrow,\downarrow}_x=\frac{\sqrt{2}(1+|v|^2)}{(|v|-1)^2}e^{-i\gamma}
(|b_{1,f,\downarrow}|^2-|b_{1,f,\uparrow}|^2).
\label{A24}
\end{eqnarray}
At this step one remains from (\ref{A20}) with 4 matching equations
(those related to $t^{f,\uparrow,\uparrow}_{\bf p}$ and $V^{d,f,\uparrow,\uparrow}_{\bf p}$ 
for $p=x,y$, i.e. 1th, 5th, 7th, 8th equalities), which must be used in 
deducing the last two unknown block operator parameters
$b_{1,f,\uparrow}$, and $b_{1,f,\downarrow}$. 

\underline{The remaining last 4 matching equations}\\
Using now the 7th and 8th equation from (\ref{A20}) and the notations
\begin{eqnarray}
&&-(V^{d,f,\uparrow,\uparrow}_x e^{-i\chi} + V^{d,f,\uparrow,\uparrow}_y) = \Theta_{\uparrow}
e^{i\delta_{\uparrow}},
\:  \Theta_{\uparrow} = |V^{d,f,\uparrow,\uparrow}_x e^{-i\chi} + V^{d,f,\uparrow,\uparrow}_y|,
\nonumber\\
&&-(V^{d,f,\uparrow,\uparrow}_x e^{-i\chi} - V^{d,f,\uparrow,\uparrow}_y) = \Theta_{\downarrow}
e^{i\delta_{\downarrow}},
\: \Theta_{\downarrow} = |V^{d,f,\uparrow,\uparrow}_x e^{-i\chi} - V^{d,f,\uparrow,\uparrow}_y|,
\label{A25}
\end{eqnarray}
one finds
\begin{eqnarray}
&&b_{1,f,\uparrow}=\sqrt{\frac{(|v|-1)^2}{2\sqrt{2}(1+|v|^2)}\frac{|v|}{|u|}}
\sqrt{\Theta_{\uparrow}} \: \:
e^{i(\gamma+\phi_u+\delta_{\uparrow})/2},
\nonumber\\
&&b_{1,f,\downarrow}=\sqrt{\frac{(|v|-1)^2}{2\sqrt{2}(1+|v|^2)}\frac{|v|}{|u|}}
\sqrt{\Theta_{\downarrow}} \: \:
e^{i(-\gamma+\phi_u+\delta_{\downarrow})/2},
\label{A26}
\end{eqnarray}
where $\chi=\phi_3+\phi_v$. Finally, from the first and 5th equation of
(\ref{A20}), the $|v|/|u|$ ratio and the 
$(\delta_{\uparrow}-\delta_{\downarrow})$ phase (with the choise $\chi=\pi/2$) 
can be expressed as
\begin{eqnarray}
\frac{|v|}{|u|}= \frac{-t^{f,\uparrow,\uparrow}_x}{\sqrt{\Theta_{\uparrow} 
\Theta_{\downarrow}}
\cos \frac{\delta_{\uparrow}-\delta_{\downarrow}}{2}}, \quad
\tan \frac{\delta_{\uparrow}-\delta_{\downarrow}}{2}=
\frac{t^{f,\uparrow,\uparrow}_y}{t^{f,\uparrow,\uparrow}_x}.
\label{A27}
\end{eqnarray}
We underline that based on the presented
solution, starting from (\ref{A26},\ref{A27}) and using 
(\ref{A12},\ref{A13},\ref{A18},\ref{A22}), all unknown block operator parameters
can be explicitly expressed in function of Hamiltonian parameters.

\section{The kinetic Hamiltonian in ${\bf k}$ space}

Transforming $\hat H_{kin}$ from (\ref{EQ3}) in ${\bf k}$ space via
$\hat c_{{\bf j},\sigma}=(1/\sqrt{N_{sit}})\sum_{\bf k}
\exp{(-i{\bf k}{\bf j})}\hat c_{{\bf k},\sigma}$, $c=d,f$, one obtains
\begin{eqnarray}
\hat H_{kin}=\sum_{\bf k} \sum_{\sigma,\sigma'} [\epsilon^{\sigma,\sigma'}_{d,{\bf k}}
\hat d^{\dagger}_{{\bf k},\sigma} \hat d_{{\bf k},\sigma'} +
\epsilon^{\sigma,\sigma'}_{f,{\bf k}} \hat f^{\dagger}_{{\bf k},\sigma} \hat f_{{\bf k},\sigma'}
+ V_{d,f,{\bf k}}^{\sigma,\sigma'} \hat d^{\dagger}_{{\bf k},\sigma} \hat f_{{\bf k},\sigma'} +
{ V_{d,f,{\bf k}}^{\sigma,\sigma'}}^* \hat f^{\dagger}_{{\bf k},\sigma} \hat d_{{\bf k},\sigma'}].
\label{B1}
\end{eqnarray}
Introducing the column vector ${\bf v}$, its transpose conjugate (the row
vector ${\bf v}^{\dagger}$), and the matrix ${\tilde W}$ as
\begin{eqnarray}
{\bf v}=\left( \begin{array}{c}
\hat d_{{\bf k},\uparrow} \\
\hat f_{{\bf k},\uparrow} \\
\hat d_{{\bf k},\downarrow} \\
\hat f_{{\bf k},\downarrow} \\
\end{array} \right), \:
{\bf v}^{\dagger}=(\hat d^{\dagger}_{{\bf k},\uparrow}, 
\hat f^{\dagger}_{{\bf k},\uparrow}, \hat d^{\dagger}_{{\bf k},\downarrow}, 
\hat f^{\dagger}_{{\bf k},\downarrow}), \:
\tilde W =
\left( \begin{array}{cccc}
\epsilon_{d,{\bf k}}^{\uparrow,\uparrow} & V_{d,f,{\bf k}}^{\uparrow,\uparrow} & 
\epsilon_{d,{\bf k}}^{\uparrow,\downarrow} & 0 \\
{V_{d,f,{\bf k}}^{\uparrow,\uparrow}}^* & \epsilon_{f,{\bf k}}^{\uparrow,\uparrow} &
0 & \epsilon_{f,{\bf k}}^{\uparrow,\downarrow}  \\
{\epsilon_{d,{\bf k}}^{\uparrow,\downarrow}}^* & 0 & \epsilon_{d,{\bf k}}^{\downarrow,
\downarrow} & V_{d,f,{\bf k}}^{\downarrow,\downarrow} \\
0 & {\epsilon_{f,{\bf k}}^{\uparrow,\downarrow}}^*  & {V_{d,f,{\bf k}}^{\downarrow,
\downarrow}}^* &  \epsilon_{f,{\bf k}}^{\downarrow,\downarrow}
\end{array} \right),
\label{B2}
\end{eqnarray}
one finds
\begin{eqnarray}
\hat H_{kin}= \sum_{\bf k} (\hat d^{\dagger}_{{\bf k},\uparrow}, \hat f^{\dagger}_{{\bf k},
\uparrow},\hat d^{\dagger}_{{\bf k},\downarrow}, \hat f^{\dagger}_{{\bf k},\downarrow} ) 
\tilde W \left( \begin{array}{c}
\hat d_{{\bf k},\uparrow} \\
\hat f_{{\bf k},\uparrow} \\
\hat d_{{\bf k},\downarrow}\\
\hat f_{{\bf k},\downarrow} \\
\end{array} \right) =
\sum_{\bf k} {\bf v}^{\dagger} \tilde W {\bf v},
\label{B3}
\end{eqnarray}
where on has for $c=d,f$ and $\sigma,\sigma'$ the expressions
\begin{eqnarray}
&&\epsilon^{\sigma,\sigma'}_{c,{\bf k}}=\epsilon^{\sigma,\sigma'}_c +
[t^{c,\sigma,\sigma'}_xe^{+i{\bf k}{\bf x}}+(t^{c,\sigma',\sigma}_x)^*
e^{-i{\bf k}{\bf x}}]+[t^{c,\sigma,\sigma'}_ye^{+i{\bf k}{\bf y}}+(t^{c,\sigma',\sigma}_y)^*
e^{-i{\bf k}{\bf y}}],
\label{B4}
\\
&&V_{d,f,{\bf k}}^{\sigma,\sigma'}=V_0^{d,f,\sigma,\sigma'}+[V^{d,f,\sigma,\sigma'}_x
e^{+i{\bf k}{\bf x}} +(V^{f,d,\sigma',\sigma}_x)^*e^{-i{\bf k}{\bf x}}]+
[V^{d,f,\sigma,\sigma'}_ye^{+i{\bf k}{\bf y}} +
(V^{f,d,\sigma',\sigma}_y)^*e^{-i{\bf k}{\bf y}}].
\nonumber
\end{eqnarray}
Note that the eigenvalue spectrum (i.e. the band structure) of $\hat H_{kin}$
is obtained from the secular equation of $\tilde W$. 

\section{Expectation values calculated with the ground state 
in ${\bf k}$ space} 

Transformed in ${\bf k}$ space based on Appendix B, the ground state 
(\ref{EQ8}), in unnormalized form becomes
\begin{eqnarray}
|\Psi_{g}\rangle= \prod_{\bf k} [\gamma_{1,{\bf k}} \hat{d}^\dagger_{{\bf k},\uparrow} 
\hat{d}^\dagger_{{\bf k},\downarrow} \hat{f}^\dagger_{{\bf k},\uparrow} + 
\gamma_{2,{\bf k}} \hat{d}^\dagger_{{\bf k},\uparrow} \hat{d}^\dagger_{{\bf k},
\downarrow} \hat{f}^\dagger_{{\bf k},\downarrow} + \gamma_{3,{\bf k}} 
\hat{d}^\dagger_{{\bf k},\uparrow} \hat{f}^\dagger_{{\bf k},\uparrow} 
\hat{f}^\dagger_{{\bf k},\downarrow} + \gamma_{4,{\bf k}} \hat{d}^\dagger_{{\bf k},
\downarrow} \hat{f}^\dagger_{{\bf k},\uparrow} \hat{f}^\dagger_{{\bf k},
\downarrow}] |0\rangle.
\label{EC1}
\end{eqnarray}
Here and hereafter, all $\sum_{\bf k}, \prod_{\bf k}$ extend over the first 
Brillouin zone. Furthermore, one has
\begin{eqnarray}
&&\gamma_{1,{\bf k}}= \gamma_\uparrow (a^*_{{\bf k},d,\uparrow} b^*_{{\bf k},d,\downarrow} 
- a^*_{{\bf k},d,\downarrow} b^*_{{\bf k},d,\uparrow}), \quad
\gamma_{2,{\bf k}}= \gamma_\downarrow (a^*_{{\bf k},d,\uparrow} b^*_{{\bf k},d,\downarrow} - 
a^*_{{\bf k},d,\downarrow} b^*_{{\bf k},d,\uparrow}),
\nonumber\\
&&\gamma_{3,{\bf k}}= \gamma_\downarrow (a^*_{{\bf k},d,\uparrow} b^*_{{\bf k},f,\uparrow} - 
a^*_{{\bf k},f,\uparrow} b^*_{{\bf k},d,\uparrow})- \gamma_\uparrow (a^*_{{\bf k},d,\uparrow} 
b^*_{{\bf k},f,\downarrow} - a^*_{{\bf k},f,\downarrow} b^*_{{\bf k},d,\uparrow}),
\nonumber\\
&&\gamma_{4,{\bf k}}= \gamma_\downarrow (a^*_{{\bf k},d,\downarrow} b^*_{{\bf k},f,\uparrow} - 
a^*_{{\bf k},f,\uparrow} b^*_{{\bf k},d,\downarrow})- \gamma_\uparrow (a^*_{{\bf k},d,\downarrow}
b^*_{{\bf k},f,\downarrow} - a^*_{{\bf k},f,\downarrow} b^*_{{\bf k},d,\downarrow}).
\label{EC2}
\end{eqnarray}
The connection between the block operator coefficients $g_{G,c,n,\alpha}$
(see (\ref{EQ6})), where g=a,b for G=A,B, furthermore c=d,f; 
$\alpha=\uparrow,\downarrow$; n=1,2,3,4; and $g_{G,{\bf k},c,\alpha}$ with g=a,b for 
G=A,B and c=d,f, is given by
\begin{eqnarray}
g_{{\bf k},c,\alpha}=g_{G,c,1,\alpha} + g_{G,c,2,\alpha} e^{-i{\bf k}{\bf x}}
+g_{G,c,3,\alpha} e^{-i{\bf k}({\bf x}+{\bf y})} + g_{G,c,4,\alpha} 
e^{-i{\bf k}{\bf y}},
\label{EC3}
\end{eqnarray}
see for notations also the first row of Appendix A, e.g.
$a_{{\bf k},d,\alpha}= a_{1,d,\alpha}+a_{2,d,\alpha} \exp (-i{\bf k}{\bf x}) +
a_{2,d,\alpha} \exp [-i{\bf k}({\bf x}+{\bf y})] + a_{4,d,\alpha}
\exp (-i{\bf k}{\bf y})$, etc.
Starting from (\ref{EC1}) the norm becomes
\begin{eqnarray}
\langle \Psi_g|\Psi_g\rangle = \prod_{\bf k} [ \sum_{n=1}^4 |\gamma_{n,
{\bf k}}|^2 ].
\label{EC4}
\end{eqnarray}
Using now the total z-spin component operator from (\ref{EQ10}), and applying
the relations
\begin{eqnarray}
&&\langle \Psi_g|\hat n^d_{{\bf k},\uparrow}|\Psi_g\rangle = |\gamma_{1,{\bf k}}|^2 +
|\gamma_{2,{\bf k}}|^2 + |\gamma_{3,{\bf k}}|^2, \quad
\langle \Psi_g|\hat n^d_{{\bf k},\downarrow}|\Psi_g\rangle = |\gamma_{1,{\bf k}}|^2 +
|\gamma_{2,{\bf k}}|^2 + |\gamma_{4,{\bf k}}|^2, 
\nonumber\\
&&\langle \Psi_g|\hat n^f_{{\bf k},\uparrow}|\Psi_g\rangle = |\gamma_{1,{\bf k}}|^2 +
|\gamma_{3,{\bf k}}|^2 + |\gamma_{4,{\bf k}}|^2, \quad
\langle \Psi_g|\hat n^f_{{\bf k},\downarrow}|\Psi_g\rangle = |\gamma_{2,{\bf k}}|^2 +
|\gamma_{3,{\bf k}}|^2 + |\gamma_{4,{\bf k}}|^2,
\label{EC5}
\end{eqnarray}
one finds 
for the ground state expectation value of $\hat S^z$ per site the expression
\begin{eqnarray}
\frac{\langle \hat S^z \rangle}{N_{sit}} = \frac{1}{2 N_{sit}} \sum_{\bf k}
\frac{|\gamma_{3,{\bf k}}|^2 -|\gamma_{4,{\bf k}}|^2 + |\gamma_{1,{\bf k}}|^2 -
|\gamma_{2,{\bf k}}|^2}{|\gamma_{1,{\bf k}}|^2 +|\gamma_{2,{\bf k}}|^2 + 
|\gamma_{3,{\bf k}}|^2 +|\gamma_{4,{\bf k}}|^2}.
\label{EC6}
\end{eqnarray}
Furthermore the ${\bf r}$ dependent long-range hopping operator
presented in (\ref{EQ11}) transformed in ${\bf k}$ space becomes
\begin{eqnarray}
\hat \Gamma_{{\bf r},c,\sigma}=\frac{2}{N_{sit}}\sum_{\bf k} 
\cos({\bf k}{\bf r}) \hat n^c_{{\bf k},\sigma}.
\label{EC7}
\end{eqnarray}
where $\sum_{\bf k}$ runs over the first Brillouin zone. If one takes for example
$c=d$, and $\sigma=\uparrow$ in (\ref{EC7}), based on (\ref{EC5}) one
finds
\begin{eqnarray}
\Gamma_{{\bf r},d,\uparrow}=
\langle \hat \Gamma_{{\bf r},d,\uparrow} \rangle = \frac{2}{N_{sit}}\sum_{\bf k}
\cos({\bf k}{\bf r}) \frac{\sum_{\alpha=1}^3 |\gamma_{\alpha,{\bf k}}|^2}{
\sum_{\alpha=1}^4 |\gamma_{\alpha,{\bf k}}|^2}.
\label{EC8}
\end{eqnarray}
The $\sum_{\bf k}$ in (\ref{EC6},\ref{EC8}) can be effectuated in the 
thermodynamic limit by chosing the length units such to have the $V_c=1$ for
the unit cell volume. In this case $N_{sit}=V$, $V$ being the sample's volume,
and one has $(1/V) \sum_{\bf k}= \int d^2{\bf k}/(2\pi)^2$, where the integral
must be taken on the first Brillouin zone.
The calculations have been done in the case [see (\ref{EQ8})]
$\gamma = \gamma_{\uparrow} = \gamma_{\downarrow}$. At the first view this seems to
be the most unfavorable case for ferromagnetism, but in fact the choise for
$\gamma_{\uparrow},\gamma_{\downarrow}$ is equivalent in our case to a 
{
choice}
of the quantification z-axis, along which $\langle \bar S^z \rangle$ 
is calculated.

\end{document}